\documentclass[letterpaper,10pt,journal,comsoc]{IEEEtran}

\usepackage{amsmath}
\allowdisplaybreaks

\usepackage{cuted}
\usepackage{amssymb}
\usepackage{amsfonts}
\usepackage{bm}
\usepackage{graphicx}
\usepackage{caption}
\usepackage{subcaption}
\usepackage{booktabs}
\usepackage{multirow}
\usepackage{array}
\usepackage{url}
\usepackage{hyperref}
\usepackage{xcolor}
\usepackage{setspace}
\usepackage{enumitem}
\usepackage{algorithm}
\usepackage{algpseudocode}
\usepackage{cite}
\usepackage{epsfig}
\usepackage{epstopdf}
\usepackage{balance}
\usepackage{comment}
\usepackage{svg}
\svgsetup{inkscapelatex=true}

\graphicspath{{Images/}}
\svgpath{{Images/}}

\IEEEoverridecommandlockouts

\makeatletter
\newcommand{\InBlock}{\hspace*{\ALG@thistlm}}
\newcommand{\BlockText}[1]{\Statex \InBlock #1}
\newcommand{\BlockMath}[1]{\Statex \InBlock $\displaystyle #1$}
\makeatother

\title{Parametric Channel Estimation and Design for Active-RIS-Assisted Communications}

\author{
  Md.~Shahriar~Sadid, Ali~A.~Nasir, Saad~Al-Ahmadi, and Samir~Al-Ghadhban%
  \thanks{
  Md.~Shahriar~Sadid, Ali~A.~Nasir, Saad~Al-Ahmadi, and Samir~Al-Ghadhban are with the Department of Electrical Engineering,
  King~Fahd~University~of~Petroleum~and~Minerals~(KFUPM), Dhahran~31261, Saudi~Arabia.
  They are also affiliated with the Center for Communication Systems and Sensing at KFUPM.
  This work was supported by KFUPM, IRC-CSS Center, under project number INCS2406.
  (e-mail: \texttt{g202313670@kfupm.edu.sa, anasir@kfupm.edu.sa, saadbd@kfupm.edu.sa, samir@kfupm.edu.sa}).
  }
}

\begin{document}
\setlength{\abovedisplayskip}{3pt plus 1pt minus 2pt}
\setlength{\belowdisplayskip}{3pt plus 1pt minus 2pt}
\markboth{}%
{Shell \MakeLowercase{\textit{et al.}}: Bare Demo of IEEEtran.cls for IEEE Journals}

\maketitle

\begin{abstract}
Reconfigurable Intelligent Surface (RIS) technology has emerged as a key enabler for future wireless communications. However, its potential is constrained by the difficulty of acquiring accurate user-to-RIS channel state information (CSI), due to the cascaded channel structure and the high pilot overhead of non-parametric methods. While both passive and active RISs (ARISs) suffer from severe multiplicative path loss, an ARIS directly compensates for this attenuation by amplifying the reflected signal, thus improving its practicality in real deployments. In this letter, we propose a parametric channel estimation method tailored for ARISs. The proposed approach integrates an ARIS model with an adaptive Maximum Likelihood Estimator (MLE) to recover the main channel parameters using a minimal number of pilots. To further enhance performance, an adaptive ARIS configuration strategy is employed, which refines the beam direction based on an initial user location estimate. Moreover, an orthogonal angle-pair codebook is used instead of the conventional Discrete Fourier Transform (DFT) codebook, significantly reducing the codebook size and ensuring reliable operation for both far-field (FF) and near-field (NF) users. Extensive simulations demonstrate that the proposed method achieves near-optimal performance with very few pilots compared to non-parametric approaches. Its performance is also benchmarked against that of a traditional passive RIS under the same total power budget to ensure fairness. Results show that the ARIS yields higher spectral efficiency (SE) by mitigating the multiplicative fading inherent in passive RISs and allocating more resources to data transmission.
\end{abstract}


\section{Introduction}
RISs enable wireless channel reconfiguration through programmable metasurfaces \cite{Zhang_2023, elmossallamy2020reconfigurable}. However, passive RISs suffer from severe double-fading in the cascaded link \cite{article2}, motivating ARISs, which amplify and reflect the incident wave.

Despite these advantages, CSI acquisition remains the main bottleneck in RIS-aided systems. Conventional methods sweep many RIS configurations, making pilot overhead scale with the number of elements and impractical for large surfaces \cite{swindlehurst2022channel}. Mobility-aware schemes require pilots proportional to RIS size \cite{Xu_channel}, while subarray methods reduce training dimension at the cost of beamforming gain or significant overhead \cite{demir2022exploiting}. Parametric estimation methods were initially developed for FF channels \cite{He2020}. NF estimators for extra-large MIMO systems exploit polar-domain sparsity to estimate user distance and angular parameters \cite{cui2022channel}. However, they lack a unified parametric framework operating seamlessly across both NF and FF regimes.

Building on these, Haghshenas \textit{et al.} \cite{haghshenas2024parametric} proposed a unified MLE that jointly estimates the direct base station (BS)-to-user and line-of-sight (LoS) user-to-RIS channels, recovering FF angles and NF angles plus distance. With adaptive RIS configurations and wide-beam initialization, it reduces overhead and remains effective under mobility. However, the framework is designed for passive RISs and does not address ARIS-specific modeling.

Channel estimation for ARIS-assisted setups is scarcely studied. While \cite{Chen2023} analyzed the Cram\'er--Rao bound under hardware impairments and proposed a joint least-squares estimator, MLE was used in \cite{mylonopoulos2023maximum} for ARIS-assisted LoS localization. Both studies focus on accuracy rather than pilot overhead. Lee \textit{et al.} \cite{lee2023channel} used active sensors in a semi-passive RIS to reduce overhead scaling, but overlooked per-element active amplification constraints.

Motivated by these gaps, we develop a parametric channel-estimation framework for ARIS-assisted systems. The proposed MLE yields closed-form estimates in both NF and FF regimes, jointly recovering the complex channel gain and user geometry: FF angles, and NF angles plus distance. We incorporate adaptive training to refine ARIS configurations from initial estimates, improving accuracy without increasing pilot length. Numerical results show our adaptive strategy reduces pilot overhead by nearly $65\%$ compared with a non-adaptive scheme. To our knowledge, this is the first parametric estimator for ARIS hardware explicitly accounting for amplified noise across both NF and FF regimes.
\section{System and Channel Model}

We consider an uplink system comprising a single-antenna BS and a single-antenna user equipment (UE) assisted by an ARIS. No direct BS-to-UE link is assumed. The ARIS comprises $N$ elements with independently tunable amplitude and phase. The elevated, fixed BS-ARIS topology guarantees an unobstructed, rank-one LoS channel $\boldsymbol{h} \in \mathbb{C}^{N}$ \cite{wu2019intelligent,alwazani2020channel}. By adopting a two-timescale acquisition model \cite{mei2022user}, this quasi-stationary infrastructure link is periodically estimated in the background (as detailed in Section V), allowing it to be mathematically treated as known prior CSI during rapid user tracking. The ARIS-to-UE channel is $\boldsymbol{g} \triangleq [g_1,\ldots,g_N]^{\mathrm{T}} \in \mathbb{C}^{N}$, which is time-varying due to user mobility. Due to severe cascaded mmWave path-loss and the array's 3D spatial filtering, heavily attenuated NLoS clusters are mathematically negligible, making this short-distance link strictly LoS-dominant \cite{liu2023near, mylonopoulos2023maximum}.

Let the UE transmit the symbol $x \sim \mathcal{CN}(0, P_d)$. The received signal at the BS is:
\begin{equation}
y = (\boldsymbol{\Phi}^{\mathrm{T}} \mathbf{D}_h \boldsymbol{g}) x + \boldsymbol{\Phi}^{\mathrm{T}} \mathbf{D}_h \boldsymbol{v} + w,
\label{eq:rx-signal}
\end{equation}
where $\boldsymbol{\Phi} \triangleq [\Phi_1,\ldots,\Phi_N]^{\mathrm{T}}$ collects the complex ARIS coefficients.. Here $\mathbf{D}_h = \operatorname{diag}(h_1,\ldots,h_N)$, and $\Phi_n = p_n e^{\mathrm{j}\theta_n}$ is the ARIS configuration, where $\theta_n$ is the phase shift and $p_n$ is the amplification factor of the $n$th element, respectively.

Two noise components are present: the ARIS amplification noise, $\boldsymbol{v} \sim \mathcal{CN}(\mathbf{0}_N,\, \sigma_v^2 \mathbf{I}_N)$, and the receiver noise, $w \sim \mathcal{CN}(0,\,\sigma^2)$.

The spectral efficiency (SE) can be expressed as:
\begin{equation}
R = \log_{2}\!\left(
1 + \frac{ P_d \left| \boldsymbol{\Phi}^{\mathrm{T}} \mathbf{D}_h \boldsymbol{g} \right|^{2} }
{\sigma^2 + \left\| \boldsymbol{\Phi}^{\mathrm{T}} \mathbf{D}_h \right\|_{2}^{2} \sigma_v^{2} }
\right).
\end{equation}

The LoS UE-to-ARIS channel vector is modeled as $\boldsymbol{g} = \sqrt{\beta} e^{\mathrm{j}\omega} \mathbf{a}(\boldsymbol{\psi})$, where $\beta \ge 0$ is the channel gain, $\omega \in [0,2\pi)$ is the phase, and $\mathbf{a}(\boldsymbol{\psi})$ is the array response. In the FF, $\boldsymbol{\psi}=(\varphi,\phi)$ contains the azimuth and elevation angles; in the NF, $\boldsymbol{\psi}=(\varphi,\phi,r)$ adds the distance $r$ to the reference element.

For the NF, the array response vector can be written as:
\begin{equation}
\mathbf{a}(\boldsymbol{\psi})
= \begin{bmatrix}
1,\, e^{-\mathrm{j}\frac{2\pi}{\lambda} s_2},\, \ldots,\, e^{-\mathrm{j}\frac{2\pi}{\lambda} s_N}
\end{bmatrix}^{\mathrm{T}},
\end{equation}
where $s_n \triangleq r_n - r_1$ denotes the path-length difference between the $n$th element and the reference element, with $r_n$ being the distance from the UE to the $n$th ARIS element.

In the FF, the response reduces to the conventional planar-wave form \cite{bjornson2017massive}:
\begin{align}
\mathbf{a}(\boldsymbol{\psi}) = \bigg[
&1, \ldots, e^{-j \frac{2\pi}{\lambda} \left[ i(n)\Delta_{\mathrm{H}} \sin(\varphi) \cos(\phi) + j(n)\Delta_{\mathrm{V}} \sin(\phi) \right]}, \nonumber \\
&\ldots, e^{-j \frac{2\pi}{\lambda} \left[ (N_{\mathrm{H}} - 1)\Delta_{\mathrm{H}} \sin(\varphi) \cos(\phi) + (N_{\mathrm{V}} - 1)\Delta_{\mathrm{V}} \sin(\phi) \right]} \bigg]^{\mathrm{T}},
\end{align}
where $i(n)$ and $j(n)$ are the horizontal and vertical coordinates of the $n$th ARIS element, $\Delta_H$ and $\Delta_V$ are the inter-element spacings along the horizontal and vertical axes, and $N_H$ and $N_V$ denote the numbers of elements per row and column, respectively. We can express $i(n)$ and $j(n)$ in terms of $\Delta_H$ and $\Delta_V$ \cite{bjornson2017massive}:
$i(n) = \operatorname{mod}(n - 1, N_{\mathrm{H}}) \quad \text{and} \quad j(n) = \left\lfloor \frac{n - 1}{N_{\mathrm{H}}} \right\rfloor$.

\section{MLE-based Estimation for Active RIS}
During the training phase, the UE transmits a known, deterministic pilot symbol $x = \sqrt{P_p}$ with power $P_p$ over $L$ consecutive time slots. The concatenated received signal vector $\boldsymbol{y} \in \mathbb{C}^{L}$ at the BS is given by:
\begin{equation}
    \boldsymbol{y} \;=\; \big( \boldsymbol{B}\, \mathbf{D}_h \boldsymbol{g} \big)\, \sqrt{P_p} \;+\; \tilde{\boldsymbol{v}}\;+\; \boldsymbol{w},
\end{equation}

where $\boldsymbol{B} \in \mathbb{C}^{L \times N}$ is the ARIS configuration matrix over these instances. The term $\tilde{\boldsymbol{v}} \in \mathbb{C}^{L}$ represents the cascaded active amplification noise, where its $l$-th element is $\tilde{v}_l = \boldsymbol{b}_l^{\mathrm{T}} \mathbf{D}_h \boldsymbol{v}_l$, with $\boldsymbol{b}_l^{\mathrm{T}}$ being the $l$-th row of $\boldsymbol{B}$ and $\boldsymbol{v}_l \sim \mathcal{CN}(\mathbf{0}_N,\, \sigma_v^2 \mathbf{I}_N)$ representing the independent thermal noise at the ARIS elements during slot $l$, and $\boldsymbol{w} \sim \mathcal{CN}(\mathbf{0}_L,\, \sigma^2 \mathbf{I}_L)$ denotes the thermal noise at the BS. The deterministic mean is $ \boldsymbol{\mu}(\boldsymbol{g}) \;=\; \big( \boldsymbol{B}\, \mathbf{D}_h \boldsymbol{g} \big)\, \sqrt{P_p},$ and the effective noise is $  \tilde{\boldsymbol{w}} \;=\; \tilde{\boldsymbol{v}} \;+\; \boldsymbol{w}; \qquad \tilde{\boldsymbol{w}} \sim \mathcal{CN}\!\left( \boldsymbol{0},\, \boldsymbol{F} \right),$ with diagonal noise covariance:
\begin{equation}
{    \boldsymbol{F} \;=\; \mathrm{diag}\!\left( f_1, \dots, f_L \right), \quad f_l \;=\; \sigma^2 \;+\; \sigma_v^2 \| \boldsymbol{b}_l^{\mathrm{T}} \mathbf{D}_h \|_2^2 .}
\end{equation}
The maximum likelihood (ML) estimate of the channel $\boldsymbol{g}$, denoted by $\hat{\boldsymbol{g}}$, maximizes the likelihood function $f_{\boldsymbol{Y}}(\boldsymbol{y}; \boldsymbol{g})$ of the received signal $\boldsymbol{y}$, given by:
\begin{equation}
f_{\boldsymbol{Y}}(\boldsymbol{y}; \boldsymbol{g}) = \frac{1}{\pi^{L} \det(\boldsymbol{F})} \,
\exp\left[
- \big( \boldsymbol{y} - \boldsymbol{\mu}(\boldsymbol{g}) \big)^{\mathrm{H}}
\boldsymbol{F}^{-1}
\big( \boldsymbol{y} - \boldsymbol{\mu}(\boldsymbol{g}) \big)
\right].
\end{equation}

Maximizing the likelihood $f_{\boldsymbol{Y}}(\boldsymbol{y};\boldsymbol{g})$ is equivalent to minimizing the quadratic form in the exponent. Since $\boldsymbol{F}$ is Hermitian positive definite, $\boldsymbol{F}^{-1}$ is Hermitian. Therefore,
\begin{equation}
\begin{aligned}
&\bigl(\boldsymbol{y}-\boldsymbol{\mu}(\boldsymbol{g})\bigr)^{\mathrm H}
\boldsymbol{F}^{-1}
\bigl(\boldsymbol{y}-\boldsymbol{\mu}(\boldsymbol{g})\bigr)
\\
&\quad =
\boldsymbol{y}^{\mathrm H}\boldsymbol{F}^{-1}\boldsymbol{y}
-2\sqrt{P_p}\,
\Re\!\left\{
\boldsymbol{y}^{\mathrm H}\boldsymbol{F}^{-1}
\boldsymbol{B}\mathbf{D}_h\boldsymbol{g}
\right\}
\\
&\qquad
+P_p
\bigl(\boldsymbol{B}\mathbf{D}_h\boldsymbol{g}\bigr)^{\mathrm H}
\boldsymbol{F}^{-1}
\bigl(\boldsymbol{B}\mathbf{D}_h\boldsymbol{g}\bigr).
\end{aligned}
\label{eq:quadratic_expansion}
\end{equation}
The constant term $\boldsymbol{y}^{\mathrm H}\boldsymbol{F}^{-1}\boldsymbol{y}$ can be discarded as it does not depend on $\boldsymbol{g}$ and thus does not affect its estimation. Hence,
\begin{equation} 
\begin{aligned}
\hat{\boldsymbol{g}}
= \operatorname*{arg\,min}_{\boldsymbol{g}} \big[
& -2 \sqrt{P_p}\, \Re\!\left\{ \boldsymbol{y}^{\mathrm{H}} \boldsymbol{F}^{-1} \boldsymbol{B} \mathbf{D}_h \boldsymbol{g} \right\} \\
& + P_p\, (\boldsymbol{B} \mathbf{D}_h \boldsymbol{g})^{\mathrm{H}} \boldsymbol{F}^{-1} (\boldsymbol{B} \mathbf{D}_h \boldsymbol{g})
\big].
\end{aligned}
\end{equation}
Now substitute $\boldsymbol{g} = \sqrt{\beta}\, e^{\mathrm{j}\omega}\, \mathbf{a}(\boldsymbol{\psi})$:
\begin{equation} 
\begin{aligned}
\{\hat{\beta},\,\hat{\omega},\,\hat{\boldsymbol{\psi}}\}
&= \operatorname*{arg\,min}_{\substack{\beta \ge 0,\, \omega \in [0,2\pi) \\ \boldsymbol{\psi} \in \Psi}}
\Big[
P_p \beta\, \mathbf{a}(\boldsymbol{\psi})^{\mathrm H} \mathbf{D}_h^{\mathrm H} \boldsymbol{B}^{\mathrm H} \boldsymbol{F}^{-1} \boldsymbol{B} \mathbf{D}_h \mathbf{a}(\boldsymbol{\psi})\\
&\quad
- 2 \sqrt{P_p \beta}\, \Re\!\left\{ e^{\mathrm{j}\omega}\, \boldsymbol{y}^{\mathrm H} \boldsymbol{F}^{-1} \boldsymbol{B} \mathbf{D}_h \mathbf{a}(\boldsymbol{\psi}) \right\}
\Big].
\end{aligned}
\label{eq:ml_beta_omega_psi_full}
\end{equation}

\textit{1) ARIS Reflected Channel Phase ($\omega$):}
The expression in \eqref{eq:ml_beta_omega_psi_full} is minimized when the real part inside the objective is maximized. Therefore, the optimal phase $\hat{\omega}$ aligns the exponential term with the phase of the complex inner product as:
\begin{equation}\label{eq:omega_hat}
\hat{\omega} \,=\, -\,\arg\!\left\{ \boldsymbol{y}^{\mathrm H} \boldsymbol{F}^{-1} \boldsymbol{B}\mathbf{D}_h \mathbf{a}(\boldsymbol{\psi}) \right\}.
\end{equation}
Substituting $\hat{\omega}$ into \eqref{eq:ml_beta_omega_psi_full} yields: 
\begin{equation}
\begin{aligned}
\{\hat{\beta},\,\hat{\boldsymbol{\psi}}\}
&=
\operatorname*{arg\,min}_{\substack{\beta \geq 0\\ \boldsymbol{\psi}\in\Psi}}
\Bigg[
P_p\beta\,
\mathbf{a}^{\mathrm H}(\boldsymbol{\psi})
\mathbf{D}_h^{\mathrm H}
\boldsymbol{B}^{\mathrm H}
\boldsymbol{F}^{-1}
\boldsymbol{B}
\mathbf{D}_h
\mathbf{a}(\boldsymbol{\psi})
\\
&\qquad
-2\sqrt{P_p\beta}\,
\left|
\boldsymbol{y}^{\mathrm H}
\boldsymbol{F}^{-1}
\boldsymbol{B}
\mathbf{D}_h
\mathbf{a}(\boldsymbol{\psi})
\right|
\Bigg].
\end{aligned}
\label{eq:ml_beta_psi}
\end{equation}

\textit{2) ARIS Reflected Channel Amplitude ($\beta$):}
This is a quadratic function in terms of $\sqrt{\beta}$, for which the minimizer can be determined in closed form as:
\begin{equation}\label{eq:beta_hat}
\hat{\beta} \;=\;
\frac{\left| \boldsymbol{y}^{\mathrm H} \boldsymbol{F}^{-1} \boldsymbol{B} \mathbf{D}_h \mathbf{a}(\boldsymbol{\psi}) \right|^{2}}
{P_p \left( \mathbf{a}(\boldsymbol{\psi})^{\mathrm H} \mathbf{D}_h^{\mathrm H} \boldsymbol{B}^{\mathrm H} \boldsymbol{F}^{-1} \boldsymbol{B} \mathbf{D}_h \mathbf{a}(\boldsymbol{\psi}) \right)^{2}}.
\end{equation}

\textit{3) ARIS Reflected Channel Array Response Parameter ($\boldsymbol{\psi}$):}
Substituting $\hat{\beta}$ into \eqref{eq:ml_beta_psi} yields:
\begin{equation}
\hat{\boldsymbol{\psi}} = \mathop{\arg\max}\limits_{\boldsymbol{\psi} \in \Psi}
\frac{\left| \boldsymbol{y}^{\mathrm{H}} \boldsymbol{F}^{-1} \boldsymbol{B} \mathbf{D}_h \mathbf{a}(\boldsymbol{\psi}) \right|^{2}}
{\mathbf{a}(\boldsymbol{\psi})^{\mathrm{H}} \mathbf{D}_h^{\mathrm{H}} \boldsymbol{B}^{\mathrm{H}} \boldsymbol{F}^{-1} \boldsymbol{B} \mathbf{D}_h \mathbf{a}(\boldsymbol{\psi})}.
\end{equation}

\section{Adaptive Beam and Amplification Factor}

\textit{Codebook \& Search Space:}
Let $\Theta=\{\boldsymbol{\theta}_k\in\mathbb{C}^N: \|\boldsymbol{\theta}_k\|_\infty=1, k=1,\ldots,C\}$ collect phase-only beams generated from orthogonal angle pairs (applicable to the NF via the Weyl identity \cite{pizzo2020degrees}). By construction, $\boldsymbol{\theta}_i^{\mathrm{H}}\boldsymbol{\theta}_j\approx 0$ for $i\neq j$, contracting the real-time search space from an exhaustive $N\times N$ DFT matrix to $C\ll N$ nearly orthogonal configuration directions.

\textit{Wide-Beam Initialization:}
Lacking prior CSI during the initial time slots, the ARIS must uniformly illuminate the target sector to guarantee spatial coverage. We adopt two complementary wide-beam phase vectors, $\boldsymbol{\theta}_{w,1}$ and $\boldsymbol{\theta}_{w,2}$, designed via constant-modulus (i.e., unit-magnitude) beampattern synthesis~\cite[Sec.~VI]{haghshenas2024parametric}. To fully exploit the ARIS active power budget $P_{\mathrm{RIS}}$ during this blind probing phase, we apply a uniform amplification factor across all elements, yielding initial reflection matrices $\boldsymbol{\Phi}_i=\sqrt{P_{\mathrm{RIS}}/N}\, \boldsymbol{\theta}_{w,i}$ for $i \in \{1,2\}$. These form the initial two rows of the configuration matrix $\boldsymbol{B}$. Unlike these initial probes, subsequent adaptive pilots dynamically allocate power via water-filling to maximize the SNR under the same strict $P_{\mathrm{RIS}}$ constraint.

\textit{Amplitude Shaping \& Phase Alignment:}
With the estimated channel $\hat{\boldsymbol{g}}$, the ideal per-element phases required for coherent combining across the cascaded links are given by $\bar{\boldsymbol{\theta}}^{\mathrm{opt}} = \exp(-\mathrm{j}\angle(\mathbf{D}_h \hat{\boldsymbol{g}}))$, which perfectly cancels the aggregate phase $h_n \hat{g}_n$ at each element. Under the ARIS active power constraint~\cite[Cor.~3]{do2024global}, the optimal per-element amplification is computed dynamically as $p_n^{(\mathrm{est})}=E \alpha_n/(\beta_n+\gamma_n)$. Here, $\alpha_n = |\hat{g}_n||h_n|$ represents the cascaded channel amplitude, $\beta_n = |h_n|^2$ is the BS-ARIS link gain, $\gamma_n = (|\hat{g}_n|^2 P_d/\sigma_v^2 + 1)/(P_{\mathrm{RIS}}/\sigma^2)$ acts as the regularization parameter balancing signal and amplified noise, and $E=(\sum_{n=1}^{N} |\alpha_n|^2 \gamma_n / (\beta_n+\gamma_n)^2)^{-1/2}$ is the power normalization scalar. After initialization, these optimized amplification factors are updated to form the ideal unconstrained ARIS configuration $\boldsymbol{\Phi}^\star=\boldsymbol{p}^{(\mathrm{est})}\odot \bar{\boldsymbol{\theta}}^{\mathrm{opt}}$.

\textit{Adaptive Pilot Selection:}
The estimate is refined by choosing the next phase-only codeword from the available candidate set that is best aligned with the ideal active vector: $\boldsymbol{\theta}^{\text{next}} = \arg\max_{\boldsymbol{\theta}\in\boldsymbol{\Theta}} |(\boldsymbol{\Phi}^{\star})^{\mathrm{H}}\boldsymbol{\theta}|$. Crucially, to prevent redundant spatial probing and progressively shrink the search space, the chosen codeword is immediately removed from the codebook for all future iterations ($\boldsymbol{\Theta} \leftarrow \boldsymbol{\Theta} \setminus \{\boldsymbol{\theta}^{\text{next}}\}$). We then construct the next ARIS configuration as $\boldsymbol{\Phi}_{\mathrm{new}}=\boldsymbol{p}^{(\mathrm{est})}\odot \boldsymbol{\theta}^{\text{next}}$ and append it to the configuration matrix $\boldsymbol{B}$. The BS collects the corresponding measurement for this new pilot slot and recalculates $\hat{\boldsymbol{g}}$ via the MLE step. This adaptive, greedy procedure (summarized in Algorithm \ref{alg:activeRIS-MLE}) iterates dynamically, refining the beam at each step until the total pilot budget $L$ is exhausted.

\begin{algorithm}[!t]
\caption{Parametric MLE for NF/FF with Active RIS and Adaptive Pilots}
\label{alg:activeRIS-MLE}
\begin{algorithmic}[1]
\Require codebook $\Theta$ (phase-only), initial wide beams $\{\boldsymbol{\theta}_{w,1},\boldsymbol{\theta}_{w,2}\}$, slots $L$,
powers $(P_p,P_{\rm RIS})$, $\mathbf{D}_h=\mathrm{diag}(\boldsymbol{h})$

\State $\boldsymbol{\Phi}_1 \gets \sqrt{P_{\rm RIS}/N}\,\boldsymbol{\theta}_{w,1}; \quad \boldsymbol{\Phi}_2 \gets \sqrt{P_{\rm RIS}/N}\,\boldsymbol{\theta}_{w,2}; \quad \boldsymbol{B}_2 \gets [\boldsymbol{\Phi}_1, \boldsymbol{\Phi}_2]^\top$

\For{$\ell=2$ \textbf{to} $L$}
  \State \textbf{Measurement:} $\boldsymbol{y}_\ell \gets \sqrt{P_p}\,\boldsymbol{B}_\ell \mathbf{D}_h \boldsymbol{g} + \boldsymbol{v}_\ell + \boldsymbol{w}_\ell$
  \BlockText{where $\boldsymbol{v}_\ell(i)=\boldsymbol{B}_\ell(i,:)\mathbf{D}_h\boldsymbol{v}_i$}

  \State \textbf{Parametric MLE:}
  \BlockMath{
    \hat{\boldsymbol{\psi}}_\ell
    = \arg\max_{\boldsymbol{\psi}\in\Psi}
    \frac{\big|\boldsymbol{y}_\ell^{\mathrm{H}}\boldsymbol{F}^{-1}\boldsymbol{B}_\ell \mathbf{D}_h \mathbf{a}(\boldsymbol{\psi})\big|^2}
         {\mathbf{a}(\boldsymbol{\psi})^{\mathrm{H}}\mathbf{D}_h^{\mathrm{H}}\boldsymbol{B}_\ell^{\mathrm{H}}\boldsymbol{F}^{-1}\boldsymbol{B}_\ell \mathbf{D}_h \mathbf{a}(\boldsymbol{\psi})}
  }

  \State \textbf{Closed-form updates:} compute $\hat{\beta}_\ell$ \eqref{eq:beta_hat}, $\hat{\omega}_\ell$ \eqref{eq:omega_hat} and set $\hat{\boldsymbol{g}}_\ell$

  \For{$k=1$ \textbf{to} $N$}
    \State $p_k^{(\mathrm{est})} \gets E\,\alpha_k/(\beta_k+\gamma_k)$ 
  \EndFor

  \If{$\ell<L$}
    \State \textbf{Phase steering:}
    \BlockMath{\bar{\boldsymbol{\theta}}_\ell^{\rm opt}\gets \exp\!\big(-j\,\angle(\mathbf{D}_h\hat{\boldsymbol{g}}_\ell)\big),}
    \BlockMath{\boldsymbol{\Phi}_\ell^\star\gets \boldsymbol{p}^{(\mathrm{est})}\odot \bar{\boldsymbol{\theta}}_\ell^{\rm opt}.}

    \State \textbf{Pilot selection:}
    \BlockMath{\boldsymbol{\theta}_{\ell+1}\gets \arg\max_{\boldsymbol{\theta}\in\boldsymbol{\Theta}} \big|(\boldsymbol{\Phi}_\ell^\star)^{\mathrm{H}}\boldsymbol{\theta}\big|}
    \BlockMath{\boldsymbol{\Phi}_{\rm new}\gets \boldsymbol{p}^{(\mathrm{est})}\odot\boldsymbol{\theta}_{\ell+1},}
    \BlockMath{\boldsymbol{B}_{\ell+1}\gets [\boldsymbol{B}_\ell^\top, \boldsymbol{\Phi}_{\rm new}]^\top,}
    \BlockMath{\boldsymbol{\Theta}\leftarrow \boldsymbol{\Theta}\setminus\{\boldsymbol{\theta}_{\ell+1}\}.}

    \State Transmit pilot $\ell{+}1$; acquire $\boldsymbol{y}_{\ell+1}$
  \EndIf
\EndFor
\State \textbf{Output:} $\hat{\boldsymbol{g}}\gets \hat{\boldsymbol{g}}_L$
\end{algorithmic}
\end{algorithm}

\begin{figure*}[!t]
    \centering
    \begin{minipage}[t]{0.48\textwidth}
        \centering
        \includegraphics[width=\linewidth]{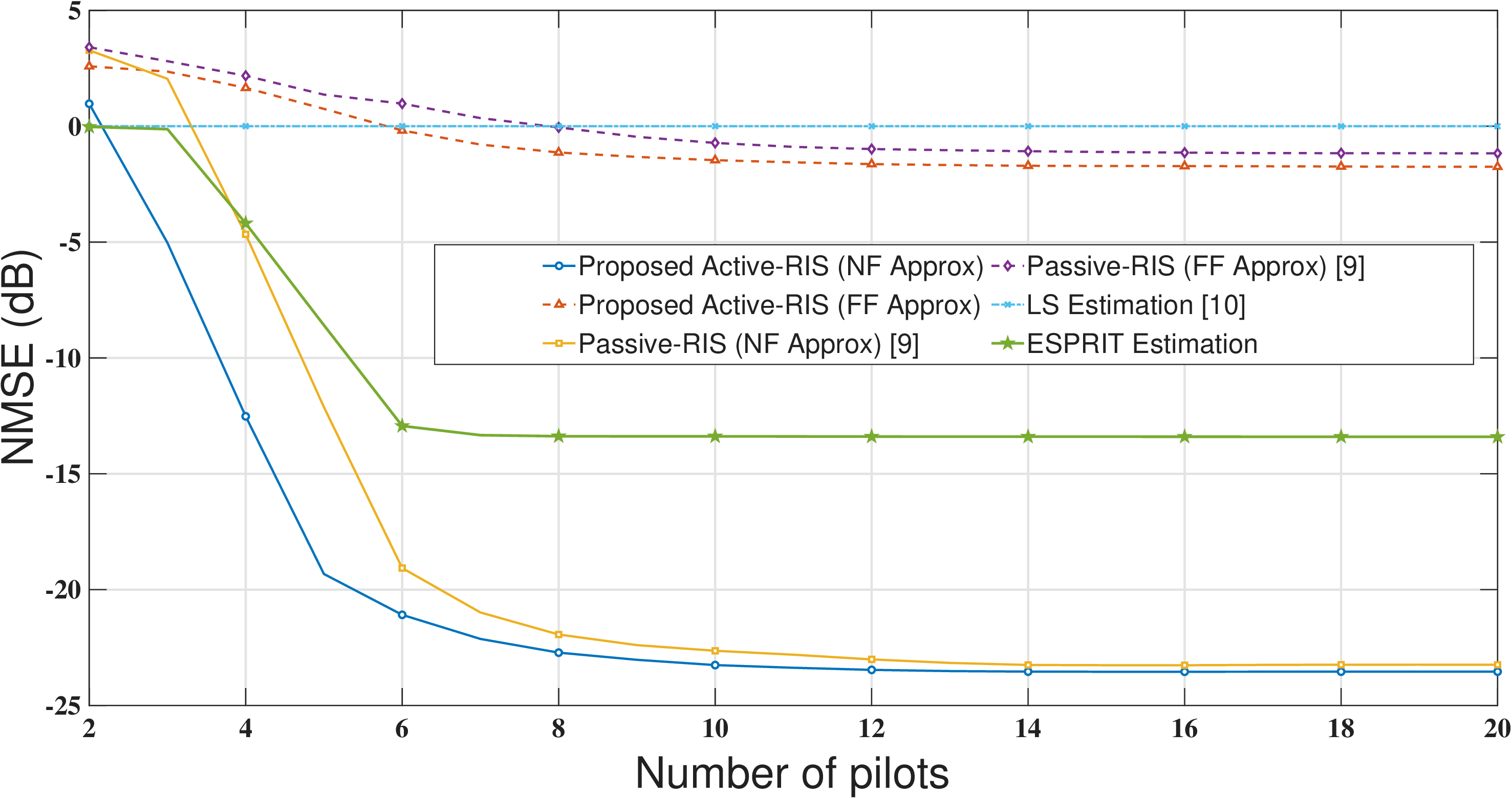}
        \caption{NMSE versus the pilot length when the user is at a random location in the NF of the ARIS.}
        \label{fig:nmse_near}
        \includegraphics[width=\linewidth]{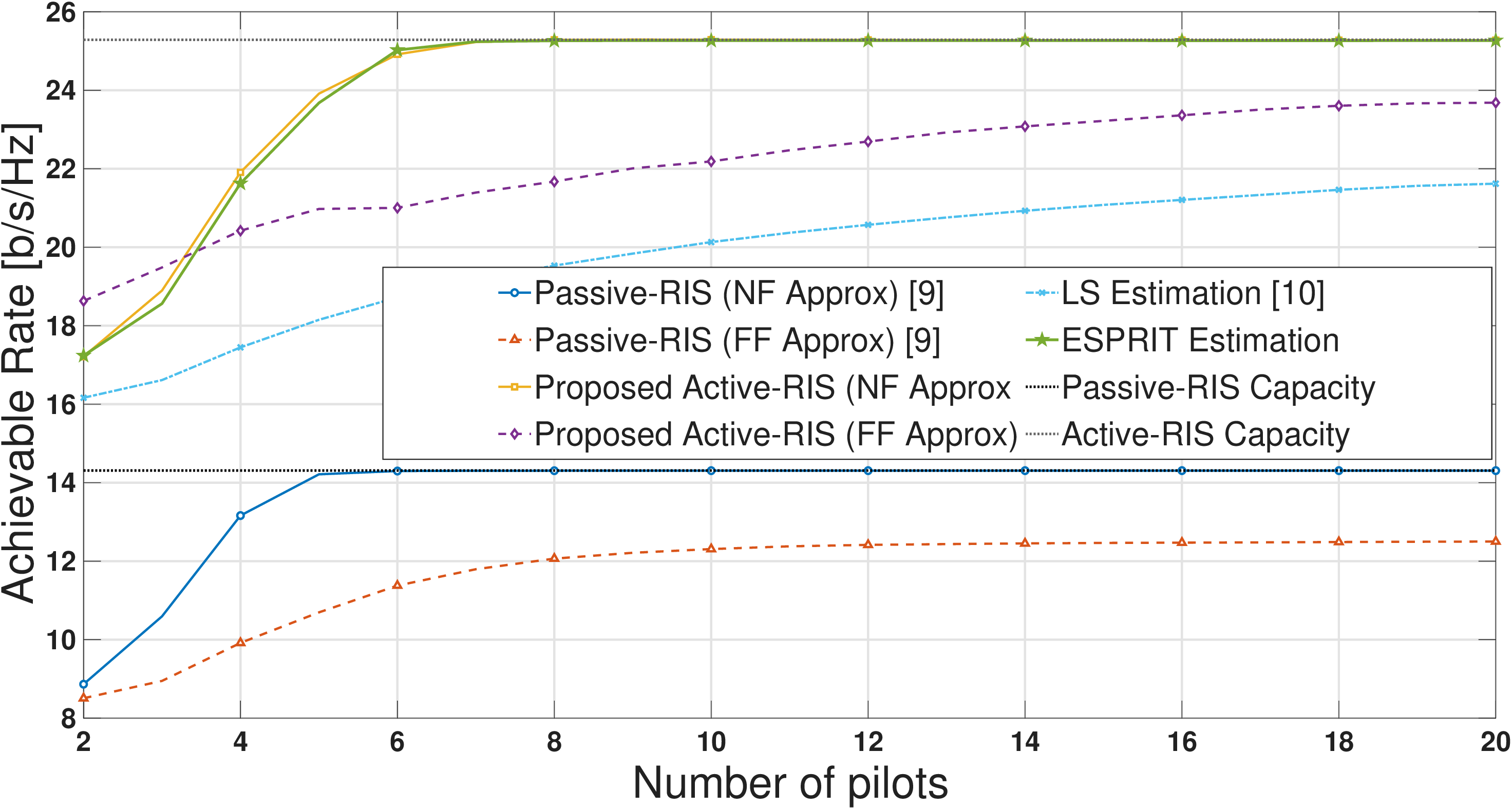}
        \caption{The average Achievable Rate versus the pilot length when the user is at a random location in the NF of the ARIS.}
        \label{fig:rate_near}
    \end{minipage}
    \hfill
    \begin{minipage}[t]{0.48\textwidth}
        \centering
        \includegraphics[width=\linewidth]{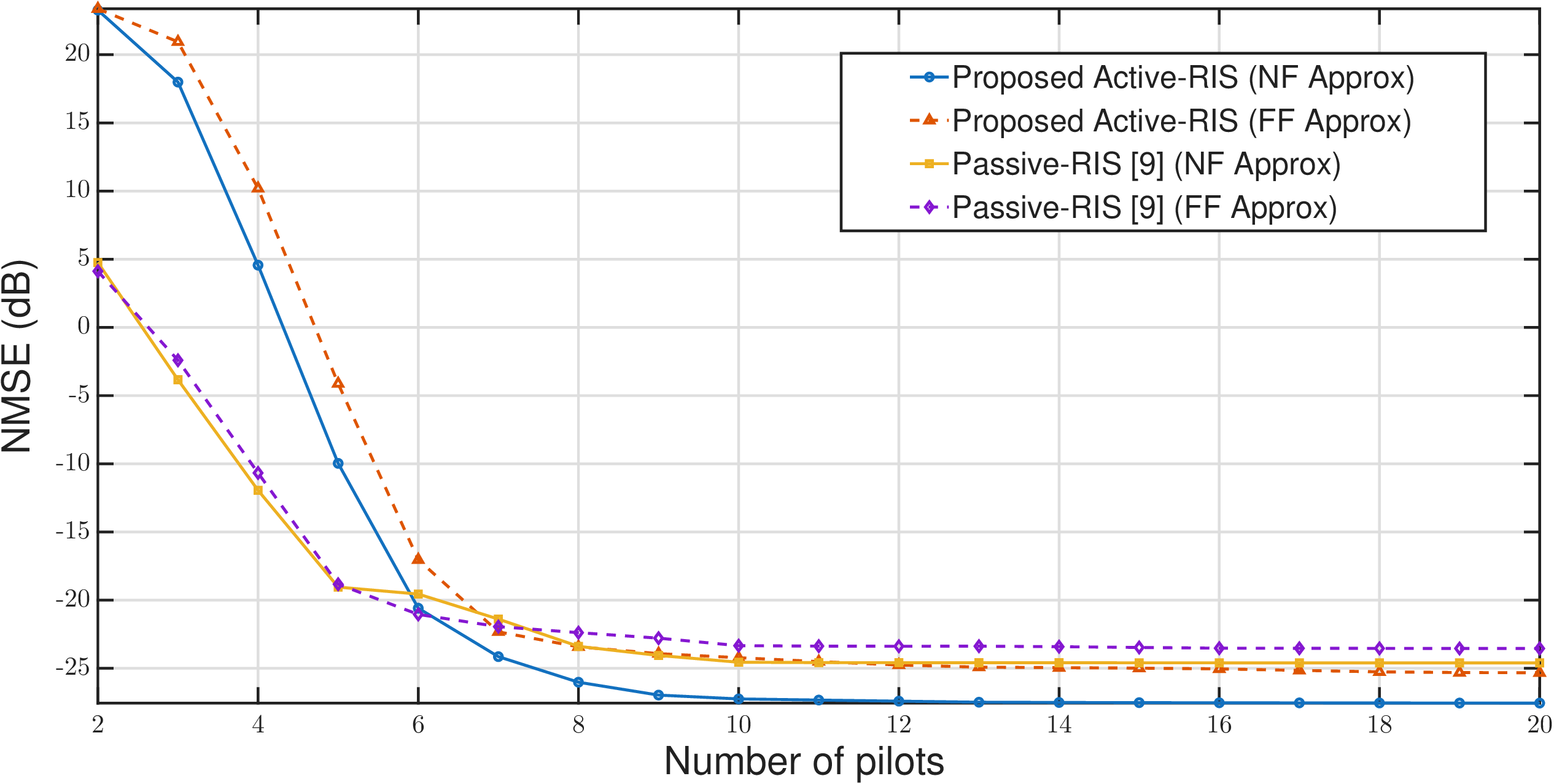}
        \caption{NMSE versus the pilot length when the user is at a random location in the FF of the ARIS.}
        \label{fig:nmse_far}
        \includegraphics[width=\linewidth]{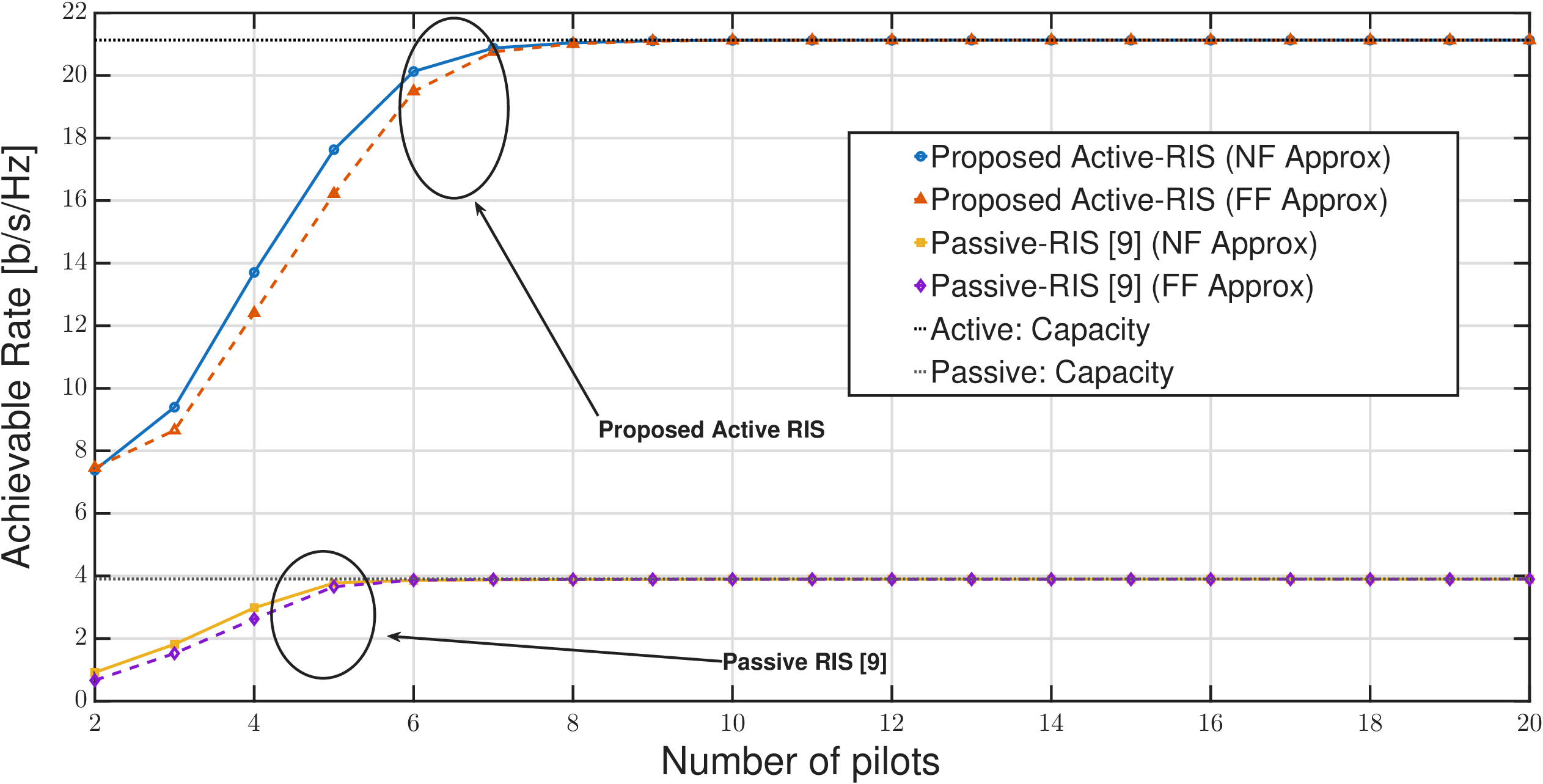}
        \caption{The average Achievable Rate versus the pilot length when the user is at a random location in the FF of the ARIS.}
        \label{fig:rate_far}
    \end{minipage}
\end{figure*}

\textit{Complexity and Signaling Overhead:}
At each iteration $\ell \le L$, evaluating the MLE requires $\mathcal{O}(\ell)$ operations for diagonal covariance matrix inversion, and $\mathcal{O}(N |\Theta_{\mathrm{rem}}|)$ operations to evaluate the remaining phase codewords, where $\Theta_{\mathrm{rem}}$ denotes the dynamically shrinking set of unselected candidate vectors ($|\Theta_{\mathrm{rem}}| \le C$). As the temporal independence of the active thermal noise ensures the covariance matrix is strictly diagonal, the standard $\mathcal{O}(L^3)$ cubic inversion bottleneck is entirely eliminated. Because the adaptive strategy converges with remarkably few pilots (e.g., $L \le 10$), the baseband processing cost is negligible for modern BS architectures. Overall, the total algorithmic complexity scales strictly as $\mathcal{O}(LNC)$, efficiently trading a marginal increase in centralized BS computation for a substantial reduction in over-the-air pilot duration.\newline
Finally, adapting the ARIS optimization between each time-slot requires real-time control signaling. Because the codebook is predefined offline, the BS strictly only transmits the integer index of the selected phase codeword alongside the dynamically calculated amplification factors. Assuming a $B$-bit quantizer for the ARIS amplifiers, this signaling payload is strictly bounded to $\lceil \log_2(C) \rceil + NB$ bits per iteration \cite{saggese2024impact}. Taking an 8-bit quantizer ($B=8$) and our setup for a massive $N=1024$ element array, and $C=813$ (the total codebook columns), this exact payload is approximately $1$ Kilobyte (KB). Because our proposed algorithm converges so rapidly (e.g., $L \le 10$ pilots), this 1 KB control signaling is executed very few times, making the cumulative overhead highly practical for real-world deployments.

\section{Simulation Results Analysis}

We evaluate the proposed estimator via Monte Carlo simulations at a carrier of $28$~GHz \cite{haghshenas2024parametric} with a $1$~MHz narrowband sub-channel.  The system has a single-antenna BS and UE assisted by an ARIS: a $32\times 32$ uniform planar array (UPA) with half-wavelength spacing ($\Delta_H=\Delta_V=\lambda/2$) placed $15$~m from the BS \cite{haghshenas2024parametric}. User angles follow $\phi,\varphi\sim\mathcal{U}[-\pi/3,\pi/3]$. NF vs.\ FF is decided by the Fraunhofer distance $d_{\mathrm f}=2D^2/\lambda$, where $D$ is the largest aperture dimension. For our setup, $d_f \approx 11$ m. For the NF, $r\sim\mathcal{U}[d_{\mathrm B},\,d_{\mathrm f}/10]$ with $d_{\mathrm B}=2D$ the Bj\"ornson distance\cite{ramezani2023near}. For the FF, $r\sim\mathcal{U}[d_{\mathrm f},\,5d_{\mathrm f}]$. Pilot SNR is taken as $10$~dB higher than the data SNR \cite{haghshenas2024parametric}.

To rigorously validate our approach, we benchmark against a passive RIS (200 mW budget) \cite{zhi2022active} -allocating 25\% power to ARIS amplifiers and 75\% to the UE for fair comparison -and the unstructured Active LS estimator \cite{Chen2023}. Furthermore, rather than assuming a perfectly known $\boldsymbol{h}$, we implement a two-timescale acquisition. Exploiting the high physical SNR inherent to the short $15$ m LoS BS-ARIS link, a minimal $2 \times 2$ ARIS sensing subarray applies 2D-ESPRIT \cite{van2002optimum} to accurately extract the 3D Angle of Arrival using merely 16 pilots. These parameters are then used to geometrically reconstruct the full 1024-element spatial channel. We evaluate our rapid user-tracking framework using this realistically estimated prior channel, $\hat{\boldsymbol{h}}$.

Figures~\ref{fig:rate_near} and~\ref{fig:rate_far} present the achievable
rates and capacity bounds for the NF and FF scenarios, respectively.
With adaptive pilot design and wide-beam initialization, the ARIS
converges within approximately $7$--$8$ pilots in both regimes and
consistently outperforms the passive RIS. In the NF, the ARIS achieves
$25.29$~bps/Hz compared with $14.31$~bps/Hz for the passive RIS,
whereas in the FF it achieves $21.01$~bps/Hz compared with
$3.87$~bps/Hz.

Figures~\ref{fig:nmse_near} and~\ref{fig:nmse_far} show the corresponding
NMSE results. In the FF, the ARIS attains a final NMSE several dB below
that of the passive RIS under both channel models. In the NF, the two
architectures exhibit comparable NMSE performance. While both models
are accurate for an FF user, the FF approximation degrades in the NF
because it neglects the user distance and therefore cannot attain the
NF capacity bound.

The Active LS and ESPRIT-based comparisons are included in the NF
results. As shown in Fig.~\ref{fig:nmse_near}, the unstructured Active
LS estimator~\cite{Chen2023} remains near $0$~dB NMSE at low pilot
overheads, whereas the proposed parametric estimator achieves much
lower NMSE with only a few pilots. Moreover,
Fig.~\ref{fig:rate_near} shows that the rate obtained using the
ESPRIT-estimated BS--ARIS channel nearly overlaps the ideal-CSI curve,
indicating that the residual ESPRIT error has negligible impact under
the considered setup.

\begin{figure}[!t]
    \centering
    \includegraphics[width=\linewidth]{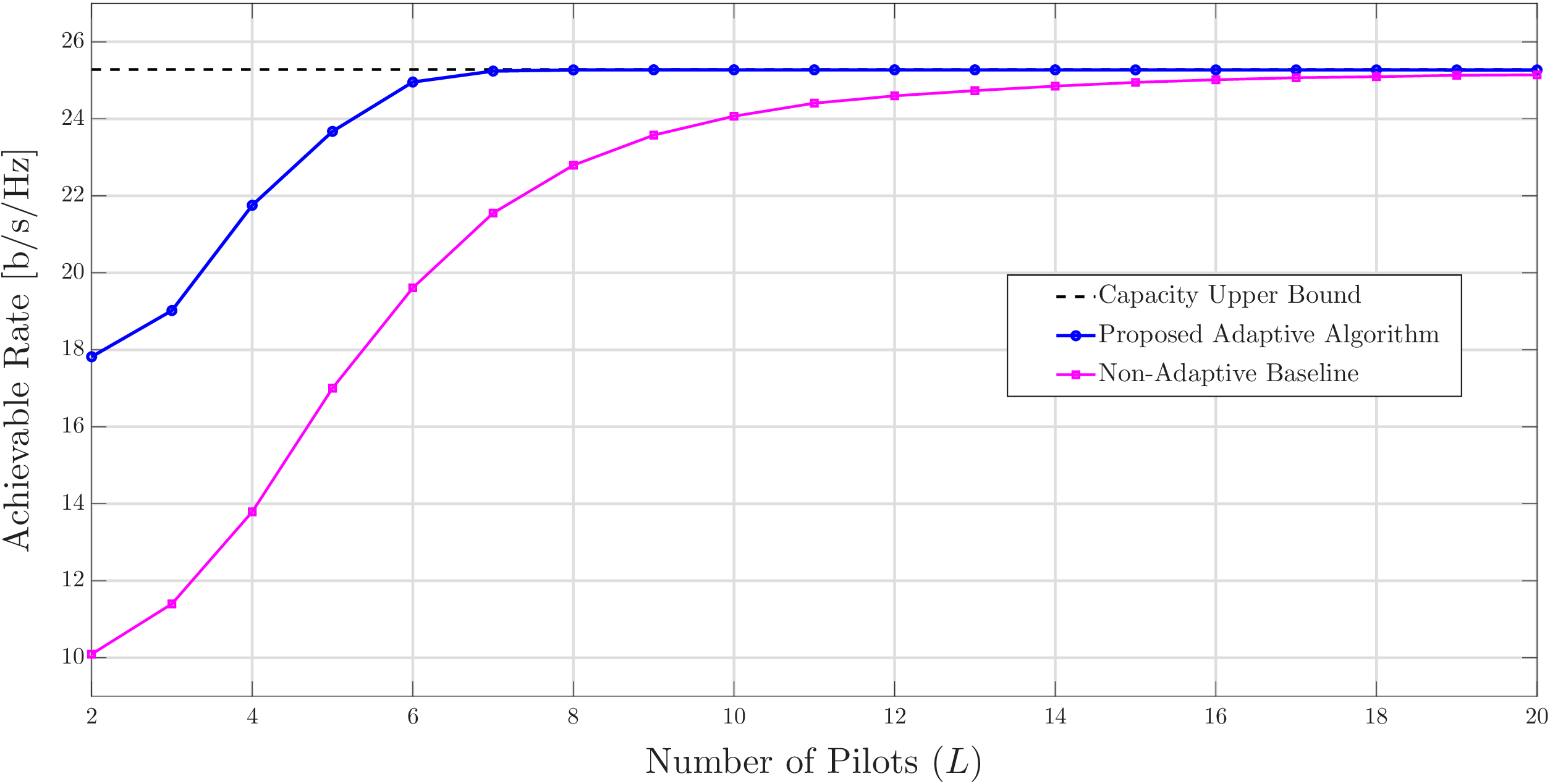}
    \caption{Comparison of the proposed algorithm with a non-adaptive method.}
    \label{fig:5}
\end{figure}
Furthermore, to isolate the algorithmic contribution of our adaptive strategy, Fig.~\ref{fig:5} evaluates the proposed method against a non-adaptive ARIS baseline. Both schemes operate under identical active hardware constraints and total power budgets. However, unlike our proposed algorithm, which dynamically recalculates the optimal ARIS state step-by-step, the non-adaptive baseline selects all phase configuration vectors uniformly at random from the predefined codebook throughout the entire training phase. As observed, while the non-adaptive baseline eventually approaches the capacity upper bound, it requires nearly $L=20$ pilots to converge. In stark contrast, our proposed adaptive tracking loop achieves optimal spectral efficiency in merely $L=7$ pilots. This explicitly confirms that the proposed adaptive framework is strictly necessary to realize the massive reduction in training overhead.

\section{Conclusion}

ARIS can enhance 6G capacity and energy efficiency by converting path loss into net link gain. Its practical use, however, depends on handling hardware limits such as nonlinearity, power constraints, and calibration drift. Semi-active RIS offers a balanced solution, while joint learning, optimization, and hardware co-design will be crucial for future gains.

\balance
\let\OLDthebibliography\thebibliography
\renewcommand\thebibliography[1]{
  \OLDthebibliography{#1}
  \setlength{\parskip}{0pt}
  \setlength{\itemsep}{0pt plus 0.3ex}
}
\bibliographystyle{IEEEtran}
\bibliography{Library}

\end{document}